\documentclass[a4paper, superscriptaddress, amsfonts, amssymb, amsmath, reprint, aps, showkeys, nofootinbib, twoside]{revtex4-2}
\usepackage[english]{babel}
\usepackage[utf8]{inputenc}
\usepackage[colorinlistoftodos, color=green!40, prependcaption]{todonotes}
\usepackage{float}
\usepackage{amsthm}
\usepackage{mathtools}
\usepackage{physics}
\usepackage{xcolor}
\usepackage{graphicx}
\usepackage[left=23mm,right=13mm,top=35mm,columnsep=15pt]{geometry} 
\usepackage{subfigure}
\usepackage{multirow}
\usepackage{adjustbox}
\usepackage{placeins}
\usepackage[T1]{fontenc}
\usepackage{lipsum}
\usepackage{csquotes}
\usepackage[pdftex, pdftitle={Article}, pdfauthor={Author}]{hyperref} 
\begin{document}
\title{Extraction of Binary Black Hole Gravitational Wave Signals from Detector Data Using Deep Learning }

\author{Chayan Chatterjee}
\email{chayan.chatterjee@research.uwa.edu.au}
\author{Linqing Wen}%
 \email{linqing.wen@uwa.edu.au}
\affiliation{%
 Department of Physics, OzGrav-UWA, The University of Western Australia,\\ 35 Stirling Hwy, Crawley, Western Australia 6009, Australia
}
 
\author{Foivos Diakogiannis}
\email{foivos.diakogiannis@icrar.org}
\affiliation{
 International Centre for Radio Astronomy Research, The University of Western Australia\\ M468, 35 Stirling Hwy, Crawley, WA, Australia
}%
 
\author{Kevin Vinsen}
\email{kevin.vinsen@icrar.org}
\affiliation{
 International Centre for Radio Astronomy Research, The University of Western Australia\\ M468, 35 Stirling Hwy, Crawley, WA, Australia
}%
 


\date{\today} 

\begin{abstract}
\noindent Accurate extractions of the detected gravitational wave (GW) signal waveforms are essential to validate a detection and to probe the astrophysics behind the sources producing the GWs. This however could be difficult in realistic scenarios where the signals detected by existing GW detectors could be contaminated with non-stationary and non-Gaussian noise. While the performance of existing waveform extraction methods are optimal, they are not fast enough for online application, which is important for multi-messenger astronomy. 
In this paper, we demonstrate that a deep learning architecture consisting of Convolutional Neural Network and bidirectional Long Short-Term Memory components can be used to extract binary black hole (BBH) GW waveforms from realistic noise in a few milli-seconds. We have  tested our network systematically on injected GW signals, with component masses uniformly distributed in the range of 10 to 80 $M_{\odot} $, on Gaussian noise and LIGO detector noise. We find that our model can extract GW waveforms with overlaps of more than 0.95 with pure Numerical Relativity templates for signals with signal-to-noise ratio (SNR) greater than six and is also robust against interfering `glitches'. We then apply our model to all ten detected BBH events from LIGO-Virgo's first (O1) and second (O2) observation runs, obtaining $\geq$ 0.97 overlaps for all ten extracted BBH waveforms with the corresponding pure templates. We discuss the implication of our result and its future applications to GW localization and mass estimation.

\end{abstract}

\keywords{Gravitational Waves, Convolutional Neural Network, Bidirectional Long Short-Term Memory, Localization, Parameter Estimation, Machine Learning, Waveform Extraction}

\maketitle



\section{\label{sec:level1}INTRODUCTION}
\noindent In its first three observation runs, the LIGO-Virgo collaboration (LVC) has made a total of 50 detections of gravitational waves (GW) from compact binary coalescence (CBC) in three observation runs \cite{GWTC2}. The detection of these signals and the subsequent analysis of their source parameters have fundamentally challenged our understanding of the origin and evolution of compact binaries in the universe. One of the most important tasks involved in GW data analysis is the extraction of pure GW waveforms from possibly non-Gaussian and non-stationary noise in which GW signals are buried at the time of detection \cite{GWTC2, BayesWave}. Having accurate estimates of the waveforms of CBC sources allow us to compare detected GW waveforms with theoretical predictions as well as make accurate predictions of the source parameters \cite{GWTC2, BayesWave}. \par
\noindent One of the main challenges involved in waveform extraction is the characterisation of the noise property of GW data, which can be non Gaussian, non-stationary and glitchy \cite{Glitch,glitch1,glitch2,glitch3,glitch4,glitch5,glitch6}.
Although noise removal is generally done in offline GW searches, it is also essential for online low-latency GW search pipelines \cite{SPIIR} to extract GW waveforms as accurately as possible. This will help in promptly validating the online detection and in accurately estimating the sky direction and chirp mass, which are essential for rapid follow-up observations \cite{em_2,em_6}. Therefore a robust and computationally efficient method is needed to extract pure waveforms from GW data. In this article, we report on the application of a deep learning model we developed to extract BBH GW signals from simulated and real detector data. \par 
\noindent Deep Learning \cite{Denoising_autoencoder_theory} is a class of Machine Learning algorithms that uses multi-layered neural networks to progressively extract features from raw input data and make predictions. The feasibility of deep learning algorithms for GW detection, parameter estimation and classification has been demonstrated previously \cite{ML_Review_paper, detection1,detection2,detection4,detection6,detection7,detection8, localization, PE1,PE2,PE3, Leila_paper, Referee_1,Referee_2,Referee_3,Referee_4,Referee_5}. The main advantage of applying deep learning to CBC sources is that these models can be pre-trained with known waveforms, prior to an observation run. In particular, when running with online searches, these models can be loaded conveniently in order to make rapid inference on live data. 
\par
\noindent The applications of deep learning models in the field of noise subtraction from image or time-series data has gained popularity over the years. The de-noising autoencoder model \cite{denoising1,denoising4}, in particular, has been found to be extremely reliable for such applications. This model consists of two parts - the encoder, which extracts essential features from noisy input data and generates a low-dimensional compressed representation of the input, and the decoder, which reconstructs the original, clean data from the compressed representation. \par  
\noindent Recently, there have been a few applications of deep learning to extract GW signal waveforms, mostly on injected GW signals. For application on GW signals injected in Gaussian noise, popular recurrent neural networks (RNN) \cite{RNN_1} like Long Short-Term Memory (LSTM) \cite{RNN_2} have been applied as encoder and decoder networks of a denoising autoencoder model and demonstrated much better signal reconstruction accuracy than traditional methods \cite{denoising1,denoising2,denoising3}. Convolutional Neural Networks (CNN) \cite{CNN_1, CNN_2} with \texttt{WaveNet}\cite{WaveNet} implementation have been used to retrieve injected BBH GW waveforms from both Gaussian and real LIGO noise \cite{denoising4}. Wei and Huerta (2019) obtained $>$ 0.97 overlaps between reconstructed signals and corresponding pure templates on GW signals injected in real noise and contaminated by simulated Gaussian and sine Gaussian glitches for SNR (defined in Section II C) > 12. They have also tested their model on injections with non-zero spin precession and have obtained similar performance. Note that the overlap (defined in Section II C) is a quantity used to denote the degree of similarity between two GW waveforms, given by their cross-correlation weighted by the noise power spectral density \cite{denoising4}. Wei and Huerta (2019) have also applied their network to one second long detector data around four detected BBH events from O1 and O2, and reported $\geq$ 0.99 overlaps for three of these events. This is the only published work that reports on deep learning applications on real detector data, at the time of writing. \par

\noindent In this paper, we have constructed, trained and tested a CNN-LSTM denoising autoencoder model on signals injected in both Gaussian and detector noise, with individual detector SNR > 8, and non-zero z-components of spins, as used by LVC's online search pipelines. Previous works involving deep learning have mostly used non-spinning waveforms for training and injection tests. Wei and Huerta (2019) have tested their model for waveform extraction on three fixed spin configurations, but have not done a detailed injection study with all possible spin combinations. In contrast, our training and injection tests consist of waveforms with non-zero z-components of spins uniformly distributed between -1 and 1. This is also the first time deep learning has been used to extract waveforms from all ten BBH events from O1 and O2, obtaining > 0.97 overlaps with templates determined by parameters described in \cite{GWTC1}
 \par

\noindent This article is organized as follows. In Section II, we describe the deep learning model we have constructed. In Section III, we demonstrate the feasibility and robustness of the method using test data. We report our results on all ten BBH events. In Section IV, we summarize our results and discuss applications and future directions of our work. The details of our model architecture and the denoising autoencoder algorithm are discussed in the Appendix.

\section{Methods}\label{Section 3}

\subsection{CNN-LSTM Model Design}
\noindent We have constructed a CNN-LSTM de-noising autoencoder model, consisting of a CNN encoder and a LSTM decoder [Fig. 1]. The details of a general denoising autoencoder model is discussed in the Appendix \cite{Denoising_autoencoder_theory}.

\begin{figure*}
  \includegraphics[width=17cm, height=10cm]{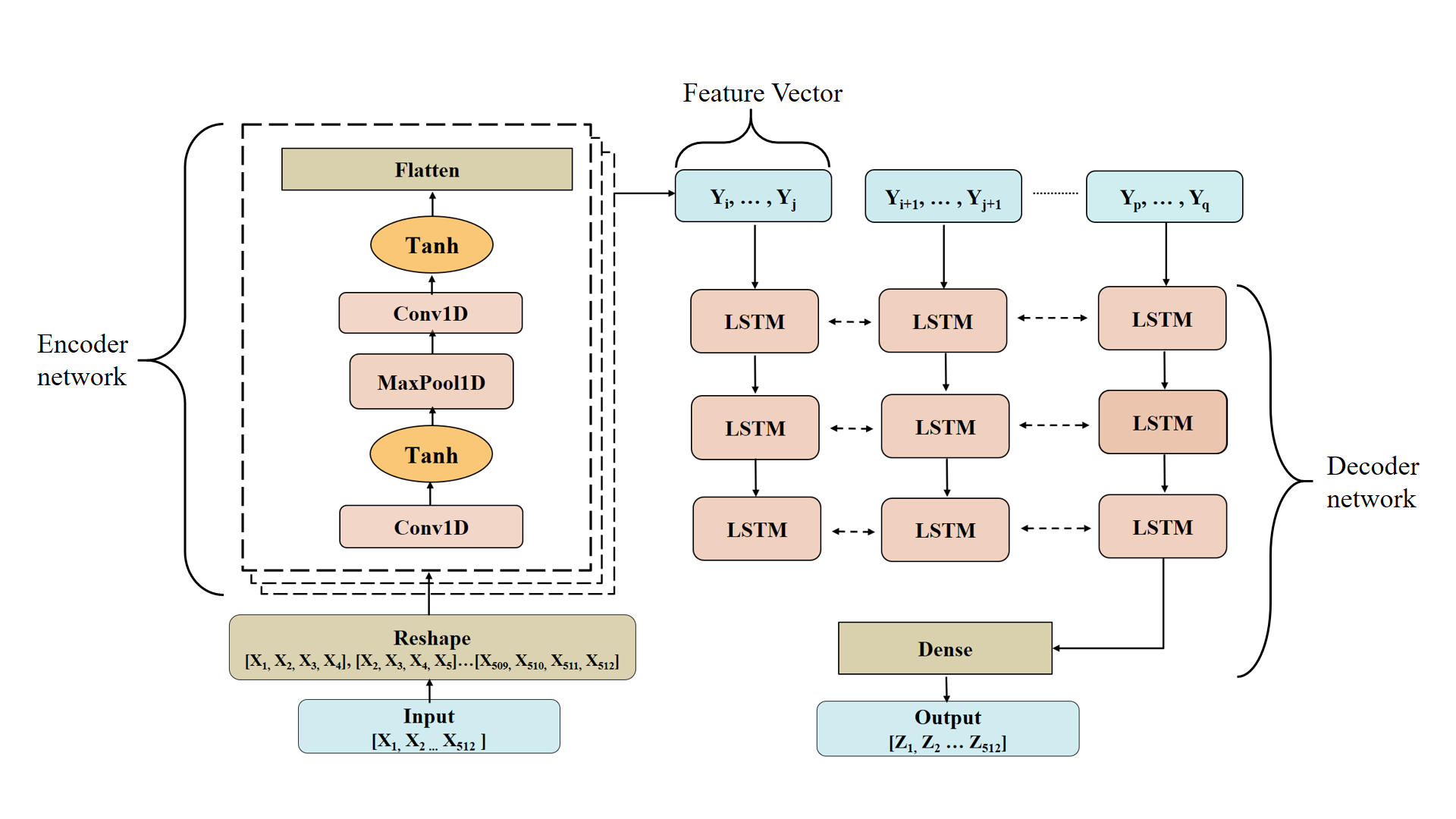}
\caption{\label{fig:2}Architecture of our CNN-LSTM denoising autoencoder model (see text). The arrows indicate flow of the data through the model. The double-headed arrows in between the LSTM layers indicate that the data is processed in both
forward and backward direction. The input data (bottom left) is first reshaped into overlapping sub-sequences and then
passed to a copy of the encoder network (boxed region). The decoder network produces the reconstructed waveforms at
the final Dense layer of the network (bottom right).}
\end{figure*}

\noindent These two networks were chosen for their known efficiency in feature extraction and time-series prediction tasks respectively. Some of the major applications of CNNs are in the fields of image classification, segmentation and time-series analysis \cite{CNN_1, CNN_2}. In CNNs, the representative patterns in data are learned using convolution kernels that scan the inputs and generate  abstracted feature maps from which the model learns to make predictions.\\
LSTMs find wide applications in the field of natural language processing and time series prediction. These networks are known to be superior in keeping track of arbitrary long-term dependencies in the input sequences and are chosen to reconstruct pure GW waveforms from the feature vector generated by the encoder network.  \par
\noindent The input data is chosen to be 0.25 sec long whitened GW strains, sampled at 2048 Hz. The amplitudes of the whitened strains and pure waveforms have very different magnitudes ($\sim 10^{2}$ for whitened strains and $ \sim 10^{-19}$ for pure waveforms). However, for training autoencoder models, the input and output data needs to be at  similar scales for efficient training. Therefore, prior to feeding the data into the model, we normalize the amplitudes of both strains and pure waveforms between the range -1 and 1. We chose this range because we found that the extracted waveforms have higher overlaps and phase consistency with pure templates when normalized between -1 and 1, rather than any other arbitrary amplitude range. After the normalization, these samples are reshaped into 516 overlapping
sub-sequences, consisting of four data points each, with
four zeroes padded at the beginning and the end to include
the full strain sample (Figure 1). The CNN encoder network
is applied to each of these 516 sub-sequences individually
for feature extraction from the underlying signal
waveform. The encoder network generates a 1-D feature
vector from each of the overlapping sub-sequences. The
LSTM network is then applied to these feature vectors
to predict the de-noised output at the next time-step
of each sub-sequence. These predictions, when pieced
together, form the extracted pure GW waveform, obtained as output at the Dense layer. \par



\noindent We experimented with different network hyper-parameters and obtained the best results with the model parameters described in Table 1 (Appendix B). The encoder part of the network consists of two one-dimensional convolutional (Conv1D) and one MaxPooling layer (MaxPool1D) \cite{CNN_1} with tan-hyperbolic activation functions. The decoder consists of three bi-directional LSTM layers with tanh activations followed by a final fully connected layer consisting of a single neuron. We have found that this rescaling leads to to better phase consistency between the extracted waveforms and pure templates, allowing the training to converge faster. \par
\noindent The Output Shape column of Table 1 refers to the shape of an individual subset of the training data, called `mini-batch', after it passes through the different layers of the network. The last column of the table shows the number of kernels used for the CNN layers and the nodes/neurons used in the LSTM and fully-connected layers. The two 1D CNN layers in the encoder consist of 32 and 16 kernels respectively, with kernel size of one. In between the two CNN layers, we use a \texttt{MaxPooling} layer to reduce the size of each sub-sequence by half. The CNN layers and the \texttt{MaxPooling} layer together extract features from the input strains which are summarized into a single vector by the \texttt{Flatten} layer. The exact copy of this encoder model is applied to each of the 516 sub-sequences of the sample simultaneously, generating a feature vector for each sub-sequence (Fig. 1). \par
\noindent The decoder consists of three bidirectional LSTM layers \cite{RNN_2}, which uses the encoded feature vector to reconstruct the pure GW waveforms. Bidirectional LSTM layers are similar to normal LSTM layers, with the exception that they process the input vector in both forward and backward directions, therefore learning from both past and future sequences of the data. This allows the model to reconstruct waveforms by effectively learning correlations between different time directions of the data segments. We use 100 neurons in each of the three LSTM layers followed by a final fully connected layer to predict a single output for each sub-sequence. The new sequence from these outputs forms the reconstructed pure GW waveform.

\subsection{Custom loss function}
\noindent A popular loss function associated with autoencoders is the mean squared error between the network predictions and the ground truths. However, for our input GW strains normalized between -1 and 1, the amplitudes away from the merger are almost close to zero and therefore the contribution to the overall mean squared error (and its gradients) from these regions is negligible compared to the error from regions near the merger peak. As a result of this, the network does not get sufficiently optimized to recover the early inspiral amplitudes with high accuracy, which affects the overlaps. Therefore, in this work, we have used a custom loss function involving two terms: the mean squared error and the negative of the Fractal Tanimoto similarity coefficient, introduced in \cite{Fractal_Tanimoto}. The Tanimoto co-efficient is a popularly used similarity metric between two vectors with values lying in the same range. We use the Fractal Tanimoto term from \cite{Fractal_Tanimoto} to reinforce amplitude and phase similarity for he entire length of the waveform considered. We use the negative of this term in order to ensure that the similarity between predictions and ground truths is maximized during the training process.

\par \noindent Specifically, the total loss function can be written as:

\begin{equation}
    L_{z,x} = \dfrac{(\textbf{x}_{i} - \textbf{z}_{i})^{2}}{n} -r_{w,z,x}^{d},
\end{equation}
where the first term is the mean squared error and \par
\begin{equation}
    r_{w,z,x}^{d} = \dfrac{\sum_{i}^{n}w_{i}.\textbf{z}_{i}. \textbf{x}_{i}}{{2^{d}{\sum_{i}^{n}w_{i}.(\textbf{z}_{i}}}^{2}+{{{\textbf{x}_{i}}}^{2}}) - (2^{d+1}-1)\sum_{i}^{n}w_{i}.\textbf{z}_{i}.\textbf{x}_{i} },
\end{equation}

\noindent is the fractal Tanimoto similarity term with added weight, ${w_{i}}$. For GWs, \textbf{z} and \textbf{x}, of dimension \emph{n} represent the autoencoder output and the clean GW waveform respectively. This term describes a similarity measure between the two vectors \textbf{z} and \textbf{x}. The effect of the $d $ parameter is to make the similarity metric steeper towards the ground truth. \cite{Fractal_Tanimoto}.\par

\noindent A comparison of the performance of the model on 3000 injections in Gaussian noise using two different loss functions: mean squared error and mean squared error - Fractal Tanimoto coefficient, is shown in Figure 2 (a) and (b). While Figure 2(a) shows around a 2\% improvement in overlaps in the range 0.90 and 1.0, on using our custom loss function over standard mean squared error for SNR between 12 and 15, we obtain, as shown in Figure 2 (b), around 12\% improvement in the same overlap range, compared to mean squared error for samples with SNR between 6 and 10. This shows that for low to moderate SNRs, our chosen custom loss function performs significantly better than mean squared error loss function.

\begin{figure*}[h]
\centering
\subfigure[]{
\includegraphics[scale=0.55]{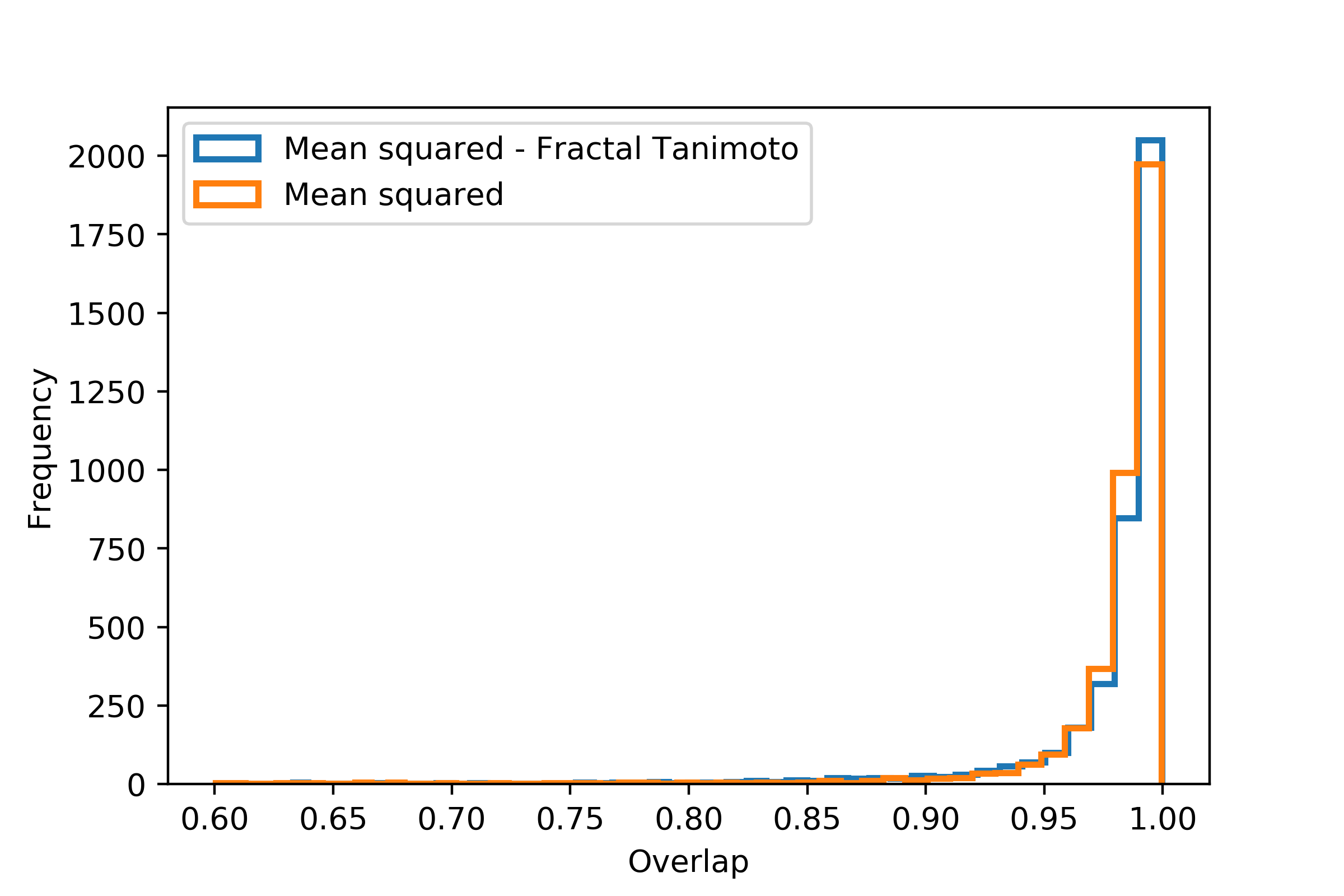}}
\subfigure[]{
\includegraphics[scale=0.55]{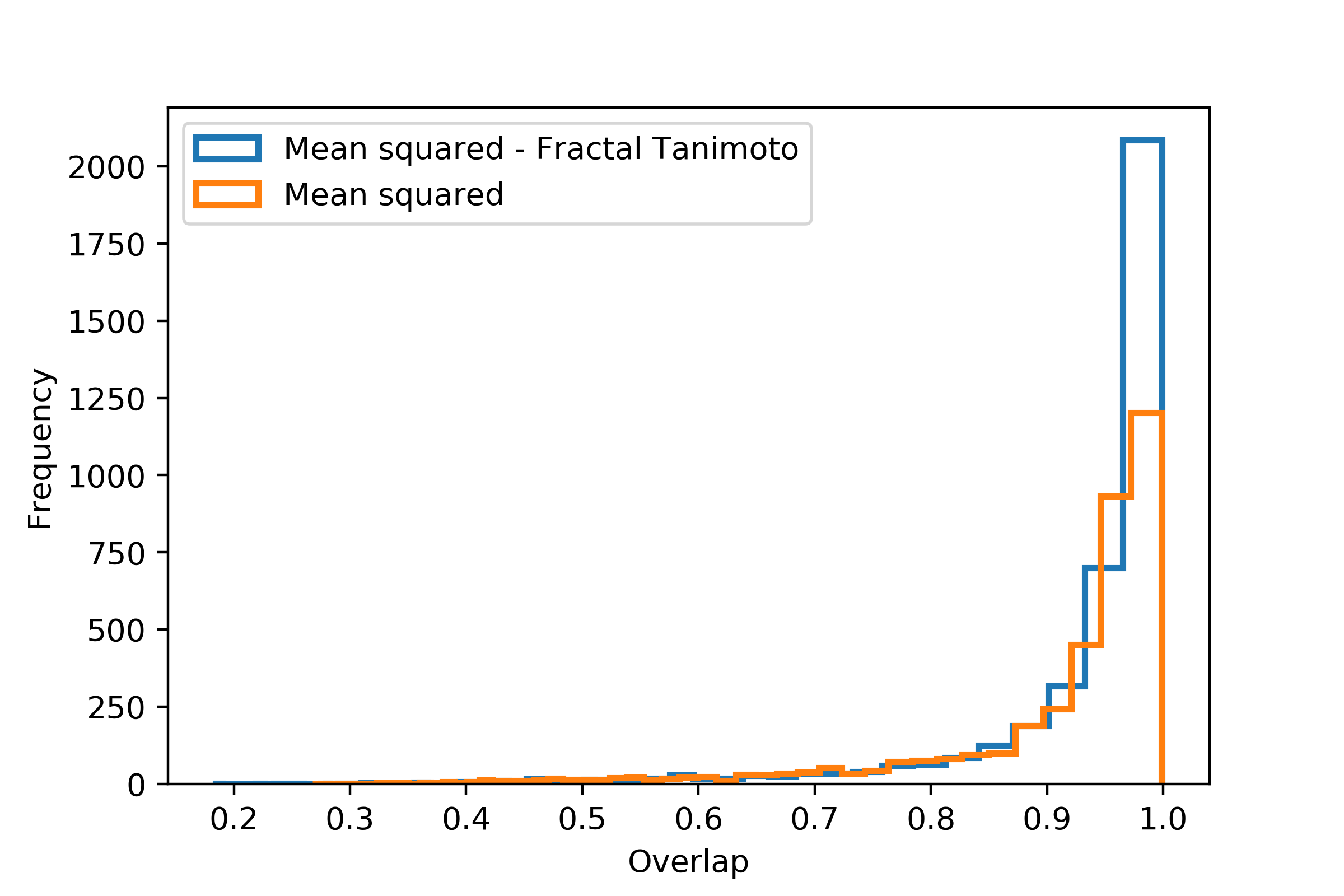}}
\caption{\label{fig:Injection_run} Histograms showing comparison of model performance for two loss functions: mean squared error and mean squared - Fractal Tanimoto term.  (a) Distribution of overlaps for samples with single detector SNR between 12 and 15. (b) Distribution of overlaps for samples with single detector SNR between 6 and 10.}
\end{figure*}

\subsection{Training and Testing}

\subsubsection{Training strategy using custom loss function}
\noindent Our training strategy follows closely the implementation described in \cite{Fractal_Tanimoto}, which involves modifying the depth parameter $d $, of the fractal Tanimoto similarity co-efficient term in the loss function, in order to boost training after the network reaches optimality. \par 

\noindent At the start of the training, we set the value of $d $ in $r_{w,z,x}^{d} $ to 0, and $w_{i} $ to 1. As the training stagnates and the loss converges after a certain number of training epochs, we reduce the learning rate and increase the value of $d $. This has the effect of `shifting gears', where the loss function is changed to a similar, but steeper one towards optimality and provides gradients that are steeper towards the ground truths (See Figure 2 in \cite{Fractal_Tanimoto}). At the beginning of training, we set the learning rate to $10^{-3}$. After the loss converges, we decrease the learning rate by a factor of ten and increased $d $ by five. Along with this, we also introduced the non-zero weight terms $w_{i} = 1/x_{i}$ in the Fractal Tanimoto coefficient. We used $d$ = 0, 5 and 10, decreasing the learning rate by a factor of 10 for each increase in $d $. In all evaluations of loss functions for $d$ > 0, the average of the loss over all $d $ values was computed:
\begin{equation}
    \langle  r_{w,z,x} \rangle^{d} = \dfrac{1}{d}\sum_{i=0}^{d-1} r_{w,z,x}^{i}
\end{equation}

\noindent In addition to this, we follow the prescription of Crum et. al. \cite{Crum} and introduce a weighting scheme in our implementation of the Fractal Tanimoto loss. During the start of training, \textbf{$w_{i}$} is set to 1, hence all data-points in the signal have the same weight. When the loss converges, we change the weights to $w_{i} = 1/x_{i} $, where $x_{i}$ are the amplitudes of the pure waveforms (ground truths), and restart training. This makes the network pay more attention to regions where the amplitude of the signal is lower, and helps improve the overall waveform extraction accuracy.

\noindent The total training time of the model is around 8.5 hrs with a P100 GPU. During testing, the model can extract pure waveforms from noise in a few milli-seconds.

\subsubsection{Sample generation}

\noindent We train and test our autoencoder with 0.25 second long GW strain data sampled at 2048 Hz. The pure BBH GW signals were simulated using the time-domain waveform approximants \texttt{SEOBNRv4} and \texttt{SpinTaylorT4}, the choice of which is based on tests described in \cite{SEOBNRv4, SpinTaylorT4}. The component masses of the black holes in our training and test sets are sampled uniformly from the range 10 to 80 solar masses. We also set uniform priors for the other intrinsic and extrinsic source parameters, including the z-component of the component spins of the black holes (-1 to 1). All of these parameters were sampled uniformly from their full range of possible values. These samples are generated using publicly available code developed by Gebhard et.al. \cite{ggwd}. The simulated waveforms were then injected into noise and the resultant strains are whitened using PyCBC's whitening method, that implements the Welch technique \cite{PyCBC}. \par
\noindent For datasets with signals injected in Gaussian noise, the advanced LIGO Zero Detuned High Power (ZDHP) PSD \cite{PSD} was used for noise generation. For injections in LIGO-Virgo detector noise, the PSD was estimated from data around the GPS times 1185579008 to 1186189312 and the noise was generated using the estimated PSD. The injected signals for the training set were generated with network optimal matched filtering signal-to-noise (SNR) ratio sampled uniformly between 10 and 25 for both Gaussian and detector noise. \par
\noindent For testing on LIGO data around detected O1 + O2 BBH events, the network was trained with injections of BBH signals on Gaussian noise (comprising 75\% of the training set) and 4096 secs detector noise right before each event (comprising 25\% of the training set), downloaded from the publicly available datasets at the Gravitational Wave Open Science Centre (GWOSC) \cite{GWOSC}. 
\noindent The neural network is trained on a Tesla K40 GPU using Python's Keras \cite{Keras} library with Tensorflow backend \cite{Tensorflow}. The model was optimized using the Adam optimizer that implements a variation of the stochastic gradient descent optimization algorithm \cite{adam}. 

\subsubsection{Overlap calculation}
\noindent In order to quantify how well the reconstructed signals match the expected waveform templates, we compute the overlap between them, which is defined as \cite{PSD}:
\begin{equation}
    \mathcal{O}(h,s) = \max_{t_{c},\phi_{c}}(\hat{h}|\hat{s}),
\end{equation} 
where, \begin{equation}
\hat{h} = h(h|h)^{-1/2},
\end{equation}
Here $\hat{h}$ and $\hat{s}$ are the normalized autoencoder output and the normalized theoretical GW template respectively. The inner product $(h|s)$ is given by:
\begin{equation} \label{eq:6}
    (h|s) = 2\int_{f_{0}}^{f_{1}}\frac{\widetilde{h}^*(f)\widetilde{s}(f) + \widetilde{h}(f)\widetilde{s}^*(f)}{S_{n}(f)}df,
\end{equation}
with $f_{0}$ = 20 Hz and $f_{1}$ = 2048 Hz. The quantities $\widetilde{h}(f)$ and $\widetilde{s}(f)$ are the Fourier transforms of $h$ and $s$ respectively and $S_{n}(f)$ is the PSD of the noise. \par

\noindent In this paper, we will characterise the performance of the autoencoder by the optimal SNRs defined as:

\begin{equation}
    \rho = \sqrt{(s|s)}
\end{equation}

and by the chirp mass, which is a key parameter for the evolution of the inspiral phase of the waveform, defined as:

\begin{equation}
    \frac{(m_{1}m_{2})^{3/5}}{(m_{1}+m_{2})^{1/5}},
\end{equation}
where $m_{1} $ and $m_{2}$ are the primary and secondary masses of the binary system.

\section{Results}\label{Section 3}

\subsection{Injection tests}
\noindent The results of our injection tests, described in this Section are shown in Figures 3 and 4.


\begin{figure*}
  \includegraphics[scale=0.4]{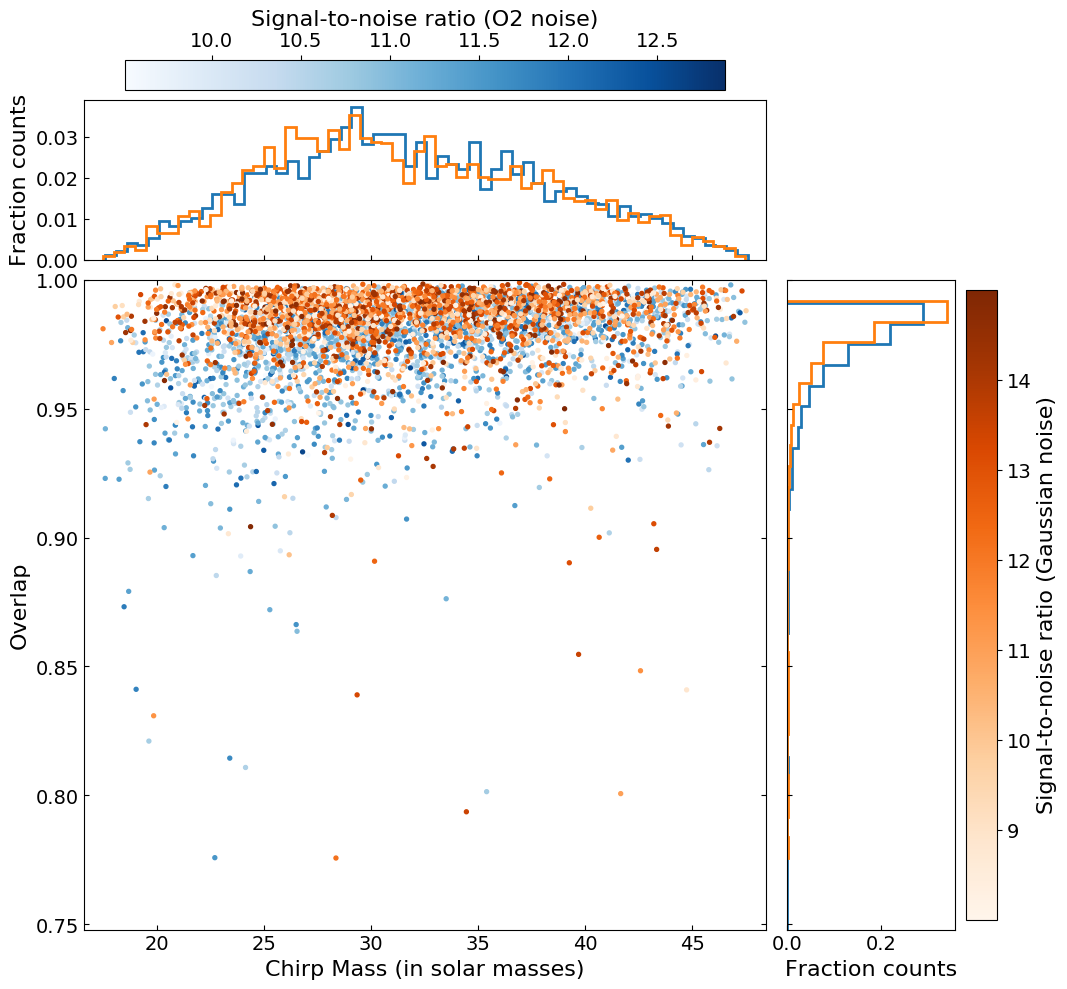}
\caption{\label{fig:Injection_run} Overlap vs. chirp mass for signals injected in Gaussian and real LIGO noise with single detector SNR between 8 and 15.}
\end{figure*}

\subsubsection{Injections in Gaussian noise}
\noindent Figure 3 shows overlaps vs. chirp mass for signals with single detector SNR between 8 and 15. We perform this test on injections in Gaussian noise, coloured by the PSD of advanced LIGO, and on injections in detector noise from O2 for the GPS times 1185579008 to 118618931. Each of these two test sets consists of 3000 samples. For Gaussian noise injections with chirp mass between 15 and 50 solar masses, we obtain overlaps of > 0.95 on more than 95\% of the samples. This shows that our model is able to learn the features of the signals buried in the noise, and also the characteristics of the noise itself, and is able to generate accurate reconstructions of the underlying pure waveforms. Similar performance is obtained on injections in detector noise (discussed in more detail in subsection 2).



\subsubsection{Injections in O2 detector noise}

\noindent For injections in detector noise, we observe in Figure 3 that our model is slightly biased against smaller chirp masses, where around 10\% of samples with chirp mass between 15 and 25 solar masses have overlaps < 0.95. For higher chirp masses, > 90\% samples have overlaps > 0.95.


\noindent In Figure 4, we demonstrate our model's performance on injections in LIGO-Virgo noise for all three detectors. We show that that while there is a wide spread of overlap for detectors with SNR $\leq$ 6, more than 90\% of samples with single detector SNR > 6 have > 0.95 overlaps with their corresponding pure templates. The overlaps are much poorer for Virgo signals, owing to the fact that the Virgo detector had very poor sensitivity during O1 and O2, and therefore had SNRs typically  lying between 0 and 5. This shows that, as expected, higher SNR leads to better overlap between pure and extracted waveforms. \par

\begin{figure*}
  \includegraphics[scale=0.5]{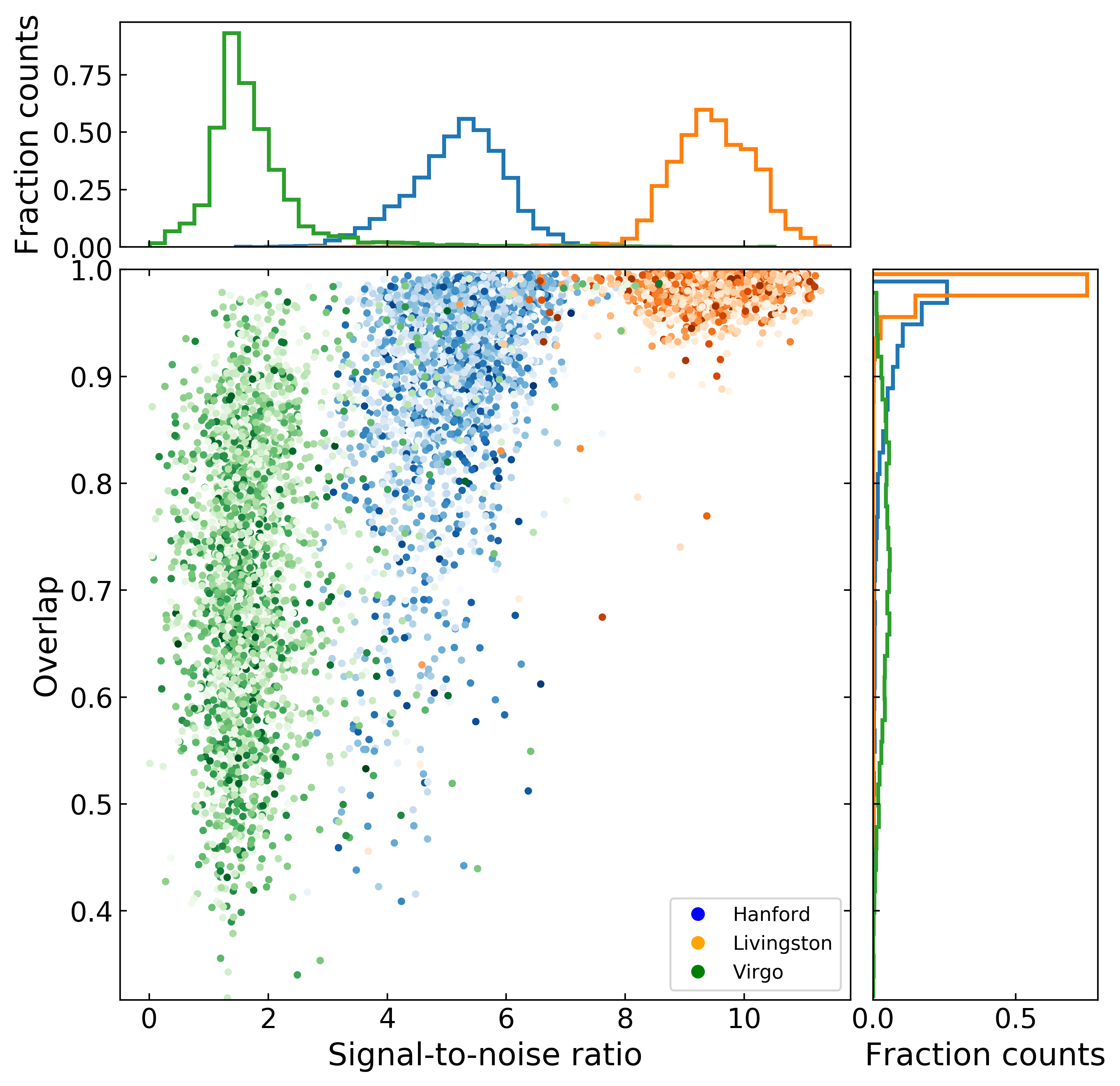}
\caption{Overlap vs. SNR for 3000 BBH samples injected in real LIGO noise from O2 for network SNR between 10 and 20. Virgo (V1) has lower SNR compared to Hanford (H1) and Livingston (L1) because the sensitivity of Virgo during O2 was much poorer than H1 and L1. The colour gradient of the scatter points for individual detectors represents the chirp mass, ranging from 20 $M_{\odot}$ to 50 $M_{\odot}$.}
\end{figure*}

\subsubsection{Robustness test of CNN-LSTM network}
\noindent We demonstrate in Figures 5-7, the robustness of our model, with two examples, (1) on pure O2 noise, with no GW injections, and (2)  on GW signals injected on O2 data but with pronounced  Gaussian glitch and sine-Gaussian glitches added to the signal. The amplitudes of the glitches and their locations within the waveform were chosen randomly by our injection algorithm. As the Figures 5 (a) and (b) and 6 (a) and (b) show, the performance of our model is unaffected by these glitches, achieving > 0.95 overlaps with pure templates for all of these cases. These glitches were chosen in particular because, if present in the data, they can often mimic the response of the detectors to an actual GW event and therefore might trigger a false detection. \par
\noindent Besides injections, we also train and test the autoencoder on additional pure O2 noise from the GPS times 1185579008 to 118618931. In response to the pure noise inputs, our model produces a straight line consistent with zero amplitude, as shown in Figure 7 (a) and (b), which is what we expect. 

\begin{figure*}
\centering
\subfigure[]{
\includegraphics[width=8.5cm, height=7cm]{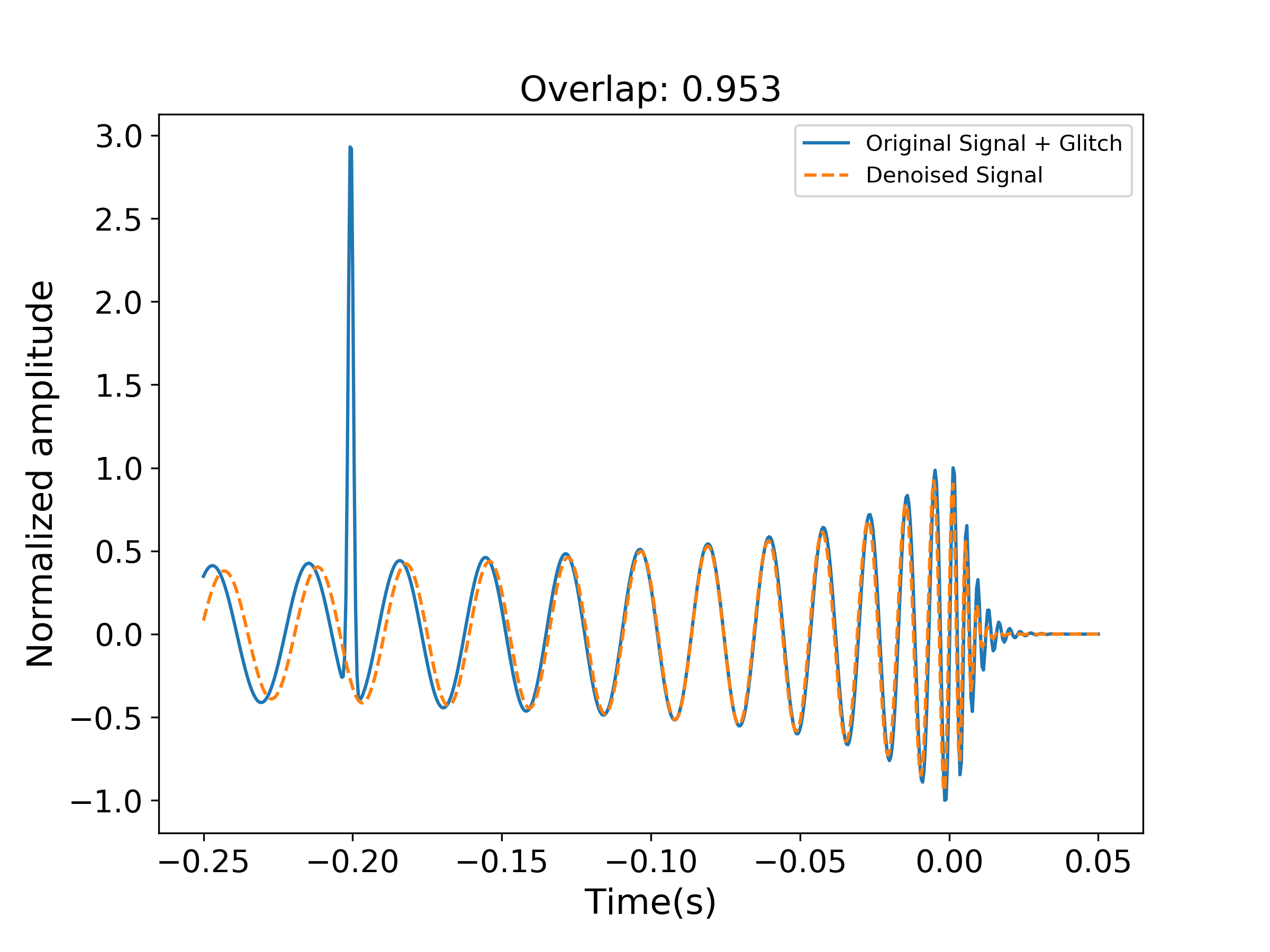}}
\subfigure[]{
\includegraphics[width=8.5cm, height=7cm]{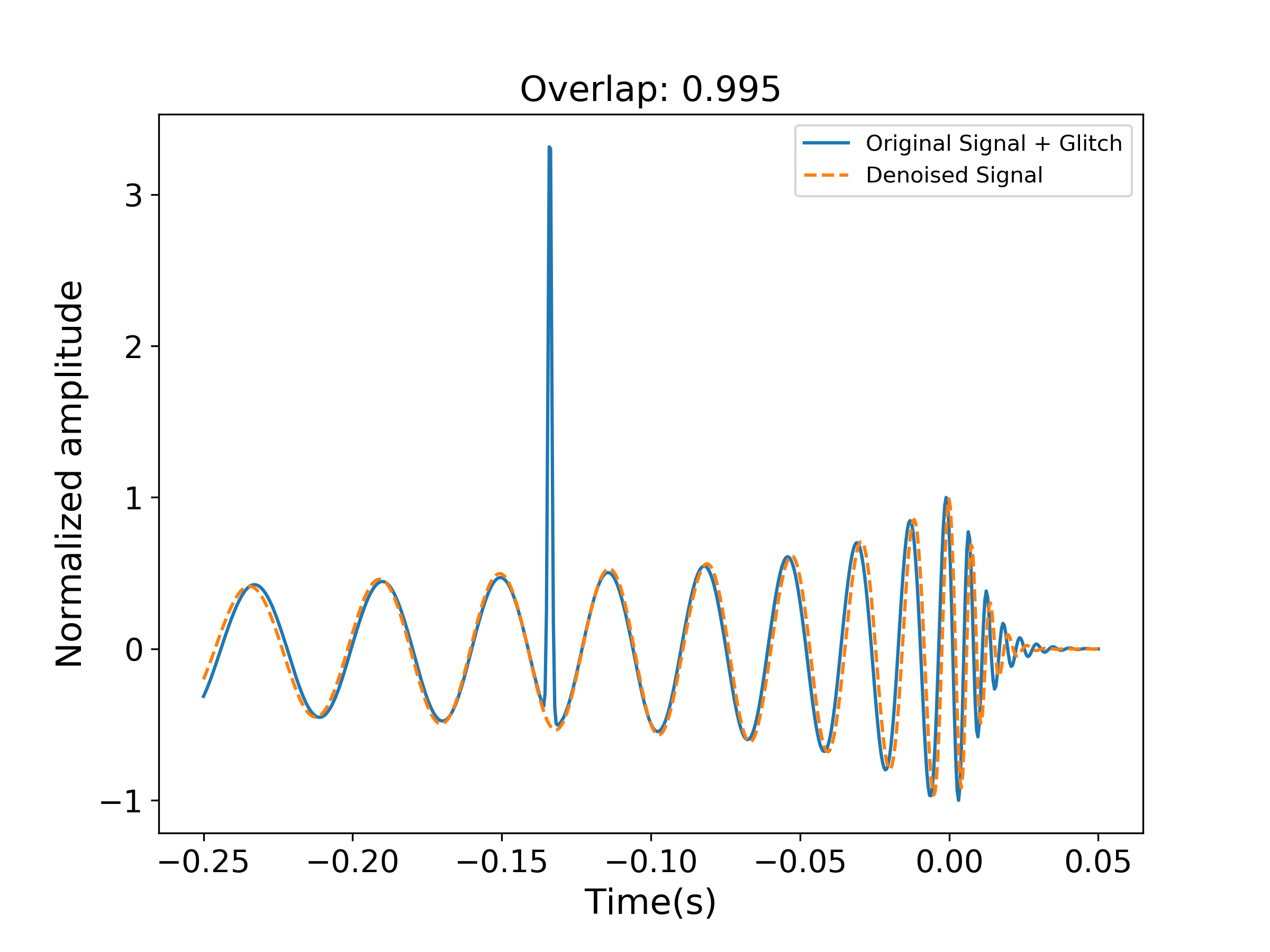}}
\caption{\label{fig:Glitch} (a) and (b) show that our network is able to remove both real noise as well as simulated Gaussian glitches that was added to noise. The glitch that was removed by the network is added to the template here for comparison.} 
\end{figure*}
\begin{figure*}
\centering
\subfigure[]{
\includegraphics[width=8.5cm, height=7cm]{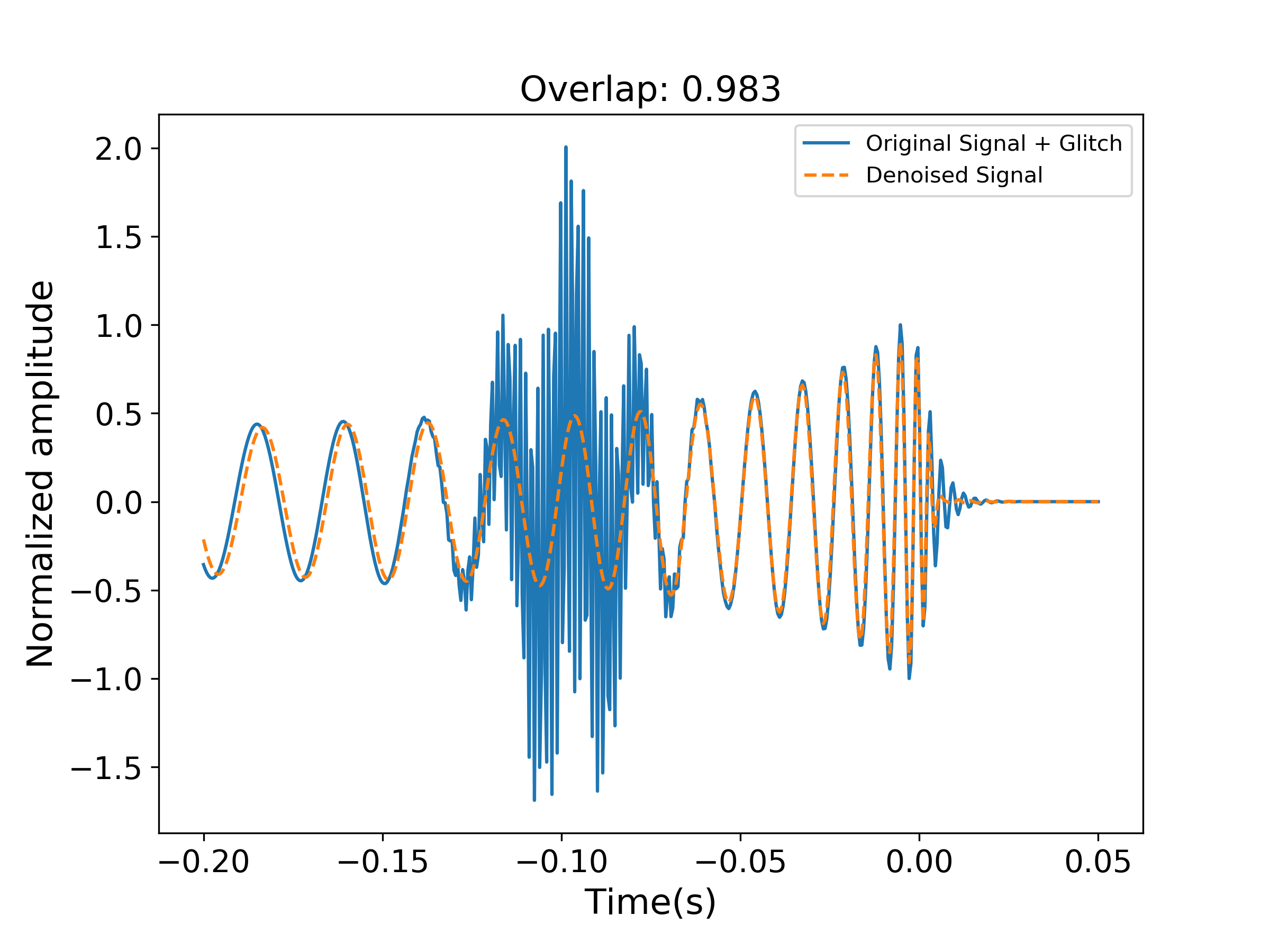}}
\subfigure[]{
\includegraphics[width=8.5cm, height=7cm]{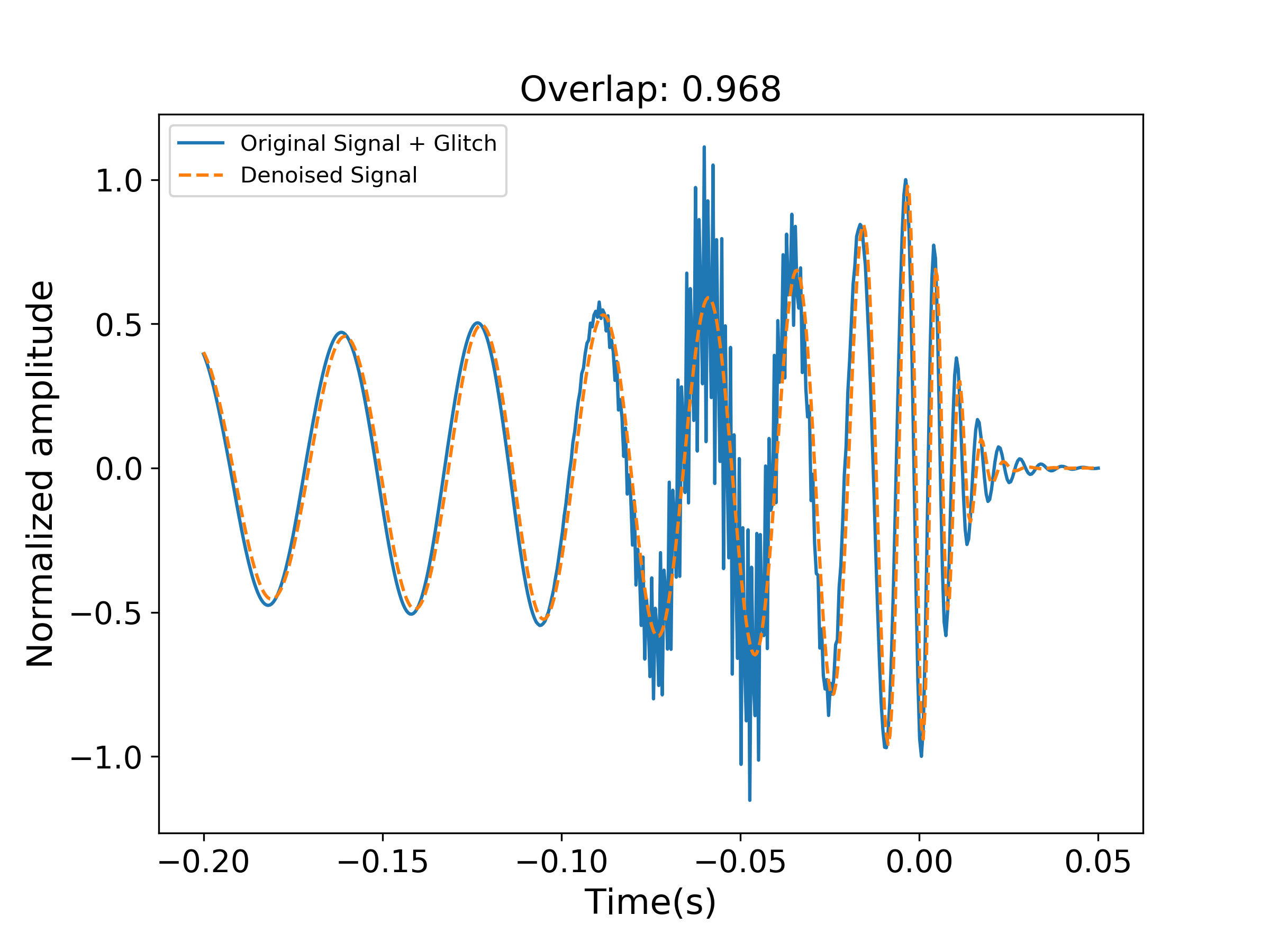}}
\caption{\label{fig:Pulse}(a) and (b) show that our network is able to remove both real noise as well as simulated sine-Gaussian glitches that was added to noise. The sine-Gaussian glitch that was removed by the network is added to the template here for comparison.}
\end{figure*}
\begin{figure*}
\centering
\subfigure[]{
\includegraphics[width=8.5cm, height=7cm]{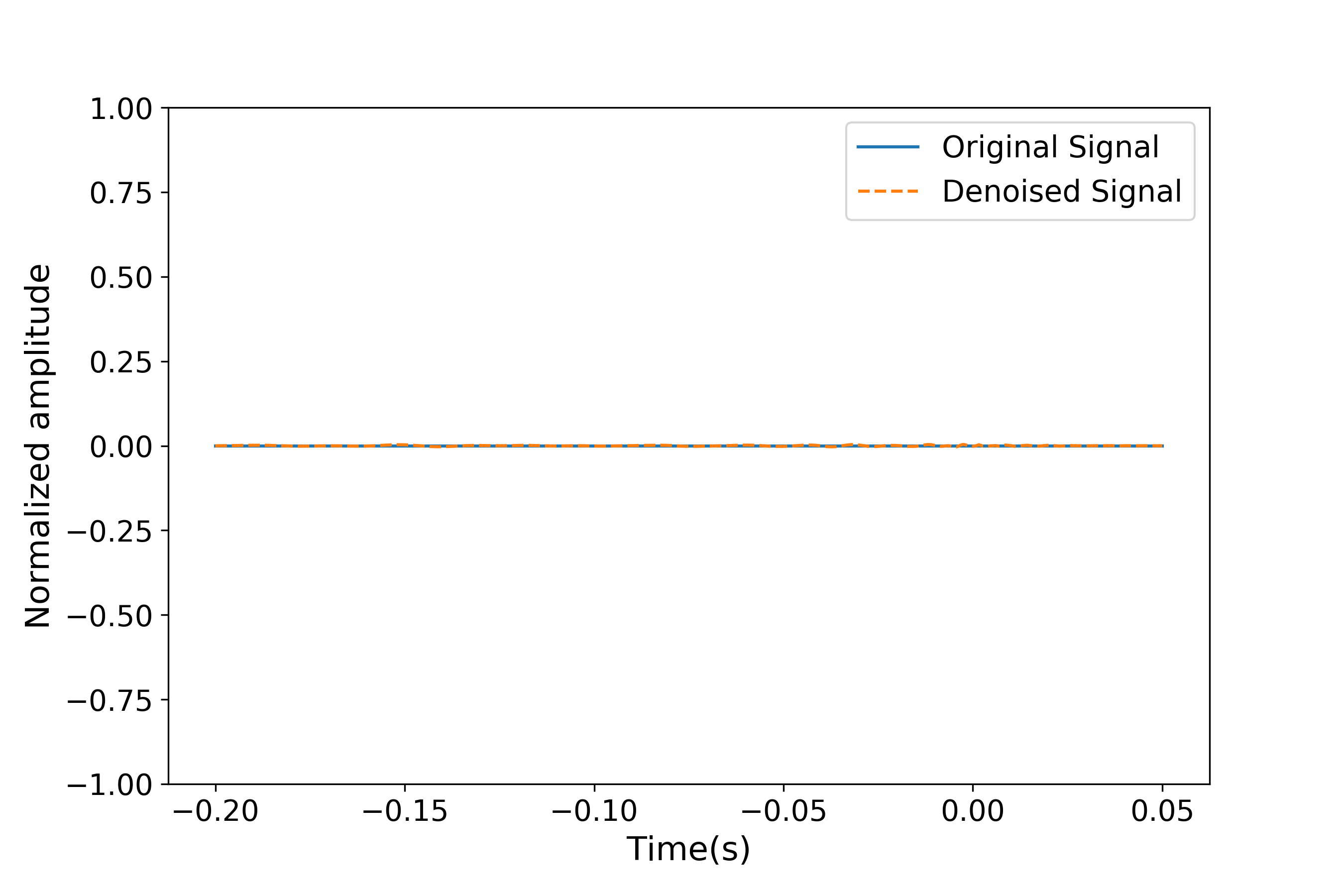}}
\subfigure[]{
\includegraphics[width=8.5cm, height=7cm]{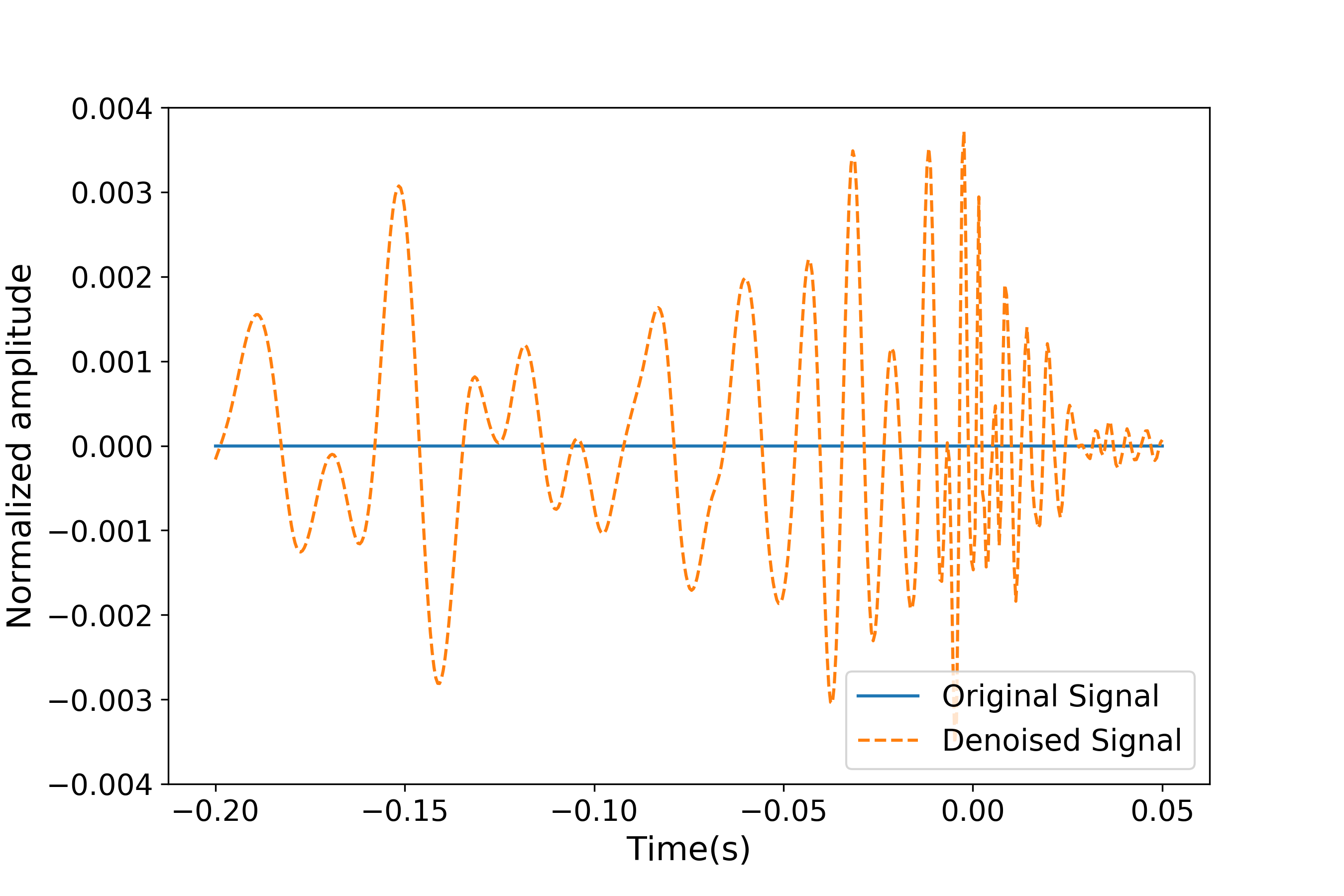}}
\caption{\label{fig:Injection_run} (a) shows that our model outputs a straight line consistent with zero amplitude when presented with only pure O2 noise with no added GW signals. (b) shows a zoomed-in version of (a).}
\end{figure*}

\subsection{Application to BBH events detected during O1 and O2}
\noindent We apply our network to all BBH events detected during O1 and O2 for waveform extraction from the detector where the events were observed with the highest SNR, as shown in Figure 8.
These events cover a large mass range, from around 7 to 55 solar masses. To obtain these results, we used the training scheme described in Section II D. We obtained $\geq$ 0.97 overlaps with the Numerical Relativity templates derived using the optimal parameters obtained from the GWTC-1 catalog \cite{GWTC1}. \par
\begin{figure*}
  \includegraphics[width=18cm, height=7cm]{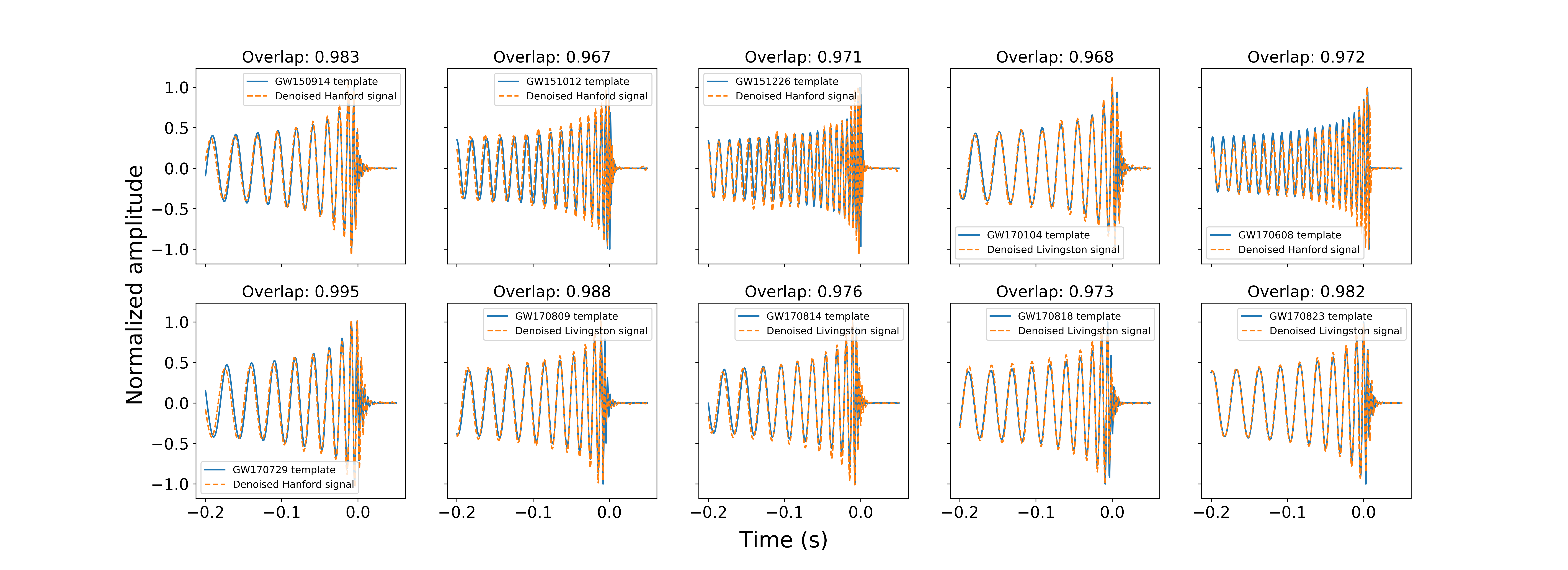}
\caption{\label{fig:2} Top panels (from left to right): Comparison between denoised signals from real data for events GW150914, GW151012, GW151226, GW170104 and GW170608 with their optimal templates. Bottom panels (from left to right): Denoised signals for events GW170729, GW170809, GW170814, GW170818, GW170823. The detectors corresponding to which the signals were observed with the largest SNR are shown here.}
\end{figure*}
\noindent We choose 0.25 sec segments of LIGO data around the merger for waveform extraction since during O1 and O2, the detectors were only sensitive to frequencies within the final 0.25 secs of the GW before the merger (see Figure 9 in \cite{GWTC1}). Since the amplitudes of the signals are quite low in the early inspiral stages of the waveform, some of the early peaks of the reconstructed signals, for example for GW150914, GW151012, GW170729 and GW170814, have worse overlaps with the pure templates compared to peaks around the merger region. \par
\noindent For GW170608, while there is good agreement on the phase, we observe a clear mismatch in the height of the peaks of the extracted waveform. This is probably because GW170608 has the smallest component masses of all the BBH events in O1 and O2, and hence has a waveform of around 3 sec in duration \cite{GWTC1}, much longer than the 0.25 secs adopted for training our model. This is consistent with our findings in Figures 3, that the extracted waveforms corresponding to low masses have worse overlaps compared to those with higher masses. \par
\noindent Wei and Huang (2019) \cite{denoising4} use their model to extract pure waveforms from 1 sec segments of LIGO data around the events GW150914, GW170104, GW170608 and GW170814. Their extracted waveforms for GW150914, GW170104 and GW170814 have > 0.99 overlaps with corresponding pure templates for  around 0.1 sec segments of the signals around the merger. We calculated the overlap over the same signal duration for the same four events and obtained similar overlaps to what is reported in Figure 8.

\section{Discussion and Conclusion}\label{Section 10}
We have designed a deep learning network of a denoising autoencoder to extract pure GW waveforms from both Gaussian as well as real LIGO-Virgo detector data. \par
\noindent We have demonstrated the robustness of this network using injections in Gaussian noise, injections in detector noise and in detector noise with modelled glitches, obtaining > 0.95 overlaps with corresponding pure signals for all of these cases. We have also reported, for the first time, waveform extraction of all BBH events from O1 and O2 and have reported > 0.97 overlaps between the denoised signals produced by our model and their corresponding optimal numerical relativity templates. As an improvement over previous work, our model is both trained and tested on GW signals from spinning black hole binaries. We have also shown that our model is able to remove structured noise anomalies that contaminate GW signals, called glitches, and produces a straight line consistent with zero amplitude when tested on pure real noise samples with no GW injections. \par
\noindent The success and reliability of our model in denoising GW strain data motivates us to apply it in conjuction with other deep learning parameter estimation models, particularly for localization (Chatterjee et. al. in preparation) and chirp mass estimation (Jacobs et. al. in preparation). These parameters rely heavily on accurate prediction of the arrival time delays, amplitudes and phases of signals in each detector, which we are able to obtain through accurate waveform extractions using our denoising autoencoder model. It takes only around 0.5 milli-seconds to extract waveforms from data using our model, with a Tesla K40 GPU. Because of its speed and accuracy this method lays the foundation for real-time mass parameter estimation and localization, with potential for implementation in online searches, which is essential for sending out Open Public Alerts (OPA) and early warning alerts for events like binary neutron star mergers. \cite{prep1, prep2}. 

\begin{acknowledgments}
\noindent This research was supported in part by the Australian
Research Council Centre of Excellence for Gravitational
Wave Discovery (OzGrav, through Project No. CE170100004). This research was undertaken with the support of computational resources from the Pople high-performance computing cluster of the Faculty of Science at the University of Western Australia.
This work used the computer resources of the OzStar computer
cluster at Swinburne University of Technology. The OzSTAR
program receives funding in part from the Astronomy
National Collaborative Research Infrastructure Strategy
(NCRIS) allocation provided by the Australian Government. This research used data obtained from the Gravitational Wave Open
Science Center (https://www.gw-openscience.org), a service of LIGO Laboratory, the LIGO Scientific Collaboration and the Virgo Collaboration. LIGO is funded by the U.S. National Science Foundation. Virgo is funded by the French Centre National de Recherche Scientifique (CNRS), the Italian Istituto Nazionale della Fisica Nucleare (INFN) and the Dutch Nikhef, with contributions by Polish and Hungarian institutes.
We would like to thank Prof. Amitava Datta and Alistair MacLeod from The University of Western Australia for their help in this work.
\end{acknowledgments}

\section{Appendix}

\subsection{De-noising Autoencoder}

\noindent In a traditional autoencoder, the encoder, given by the function $f_{\theta}(\textbf{x})$ maps the input data \textbf{x} to a hidden representation \textbf{y} through the mapping \textbf{y} = $f_{\theta}(\textbf{x})$ = $s(\textbf{Wx+b})$, where $\theta = \{\textbf{W}, \textbf{b}\}$ represents the weight matrix, \textbf{W} and bias vector, \textbf{b} and $s$ is a non-linear activation function that allows the model to learn non-linear features of the data, thereby making it more robust. The hidden vector \textbf{y} is then mapped to the reconstruction vector \textbf{z} by the decoder network, $g_{\theta'}$(\textbf{y}) through the mapping \textbf{z} = $g_{\theta'}$(\textbf{y}) = $s(\textbf{W'y + b'})$. Here $\textbf{W'}$ and $\textbf{b'}$ are the weights and biases of the decoder network. These weights and biases are optimized during the training process to minimize the reconstruction error between the autoencoder output \textbf{z} and the original input \textbf{x}. The optimization can be written as:

\begin{equation}
    \theta_{j}^{*}, \theta_{j}^{'*} \longrightarrow \theta_{k}^{*}, \theta_{k}^{'*}  = \mathop{\text{arg min}}_{\theta, \theta'} \dfrac{1}{n}\sum\limits_{i=1}^n L(\textbf{x}^{(i)},g_{\theta'}(f_{\theta}(\textbf{x}^{i}))),
\end{equation}
where $L$ is the loss function which is usually chosen to be the mean squared error between \textbf{x} and \textbf{z}, calculated over $n$ samples. $\theta_{j}^{*}, \theta_{j}^{'*} $ and $\theta_{k}^{*}, \theta_{k}^{'*}$ are the old and updated weights respectively, obtained through training, using optimization algorithms like Stochastic Gradient Descent \cite{SGD}. In the optimization process, the parameters $\theta_{j}^{*}, \theta_{j}^{'*} $ of the encoders and decoder networks are tuned to ensure the loss function, $L $ is minimized. The operator $\mathop{\text{arg min}}_{\theta, \theta'} $ denotes values of  parameters $\theta_{k}^{*}, \theta_{k}^{'*}$ that minimize $L $. 

\begin{figure}[H]
\includegraphics[width=8.5cm]{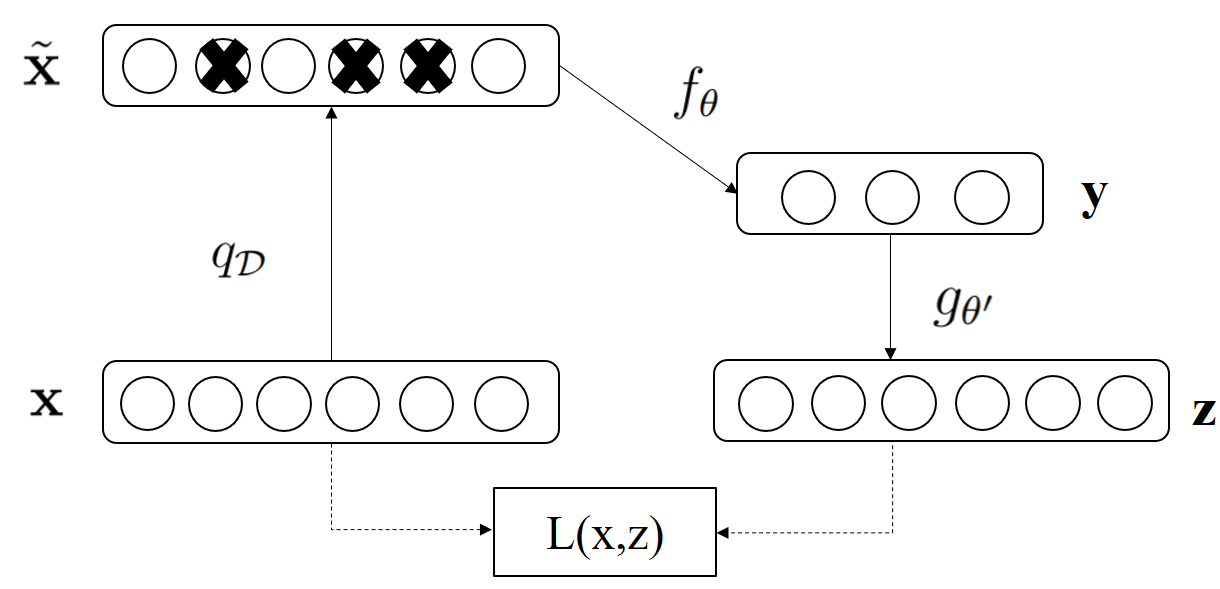}
\caption{\label{fig:DAE} Schematic diagram of a denoising autoencoder. The input \textbf{x} is corrupted to $\tilde{\textbf{x}}$. The autoencoder maps it to a hidden representation \textbf{y} and then tries to reconstruct the original input \textbf{x}.}
\end{figure}

\noindent In contrast, in a denoising autoencoder model (Figure 9), the original, clean data \textbf{x} is first corrupted by noise, to obtain the noisy inputs $\widetilde{\textbf{x}}$. The corruption process is denoted by $ q_{D}(\widetilde{\textbf{x}}|\textbf{x})$. The corrupted input $\widetilde{\textbf{x}}$ is then passed through the encoder network $f_{\theta}(\widetilde{\textbf{x}})$ = $s(\textbf{W}\widetilde{\textbf{x}}+\textbf{b})$ which maps the input to the hidden representation \textbf{y}. The decoder network $g_{\theta'}({\textbf{y}})$ = $s(\textbf{W'y+b'})$ then outputs $\textbf{z} $, the network's reconstruction of the original clean data, \textbf{x} .
The parameters $\theta$ and $\theta'$ of the model are optimized during training to minimize the reconstruction error between the autoencoder outputs $\textbf{z}$ and $\textbf{x}$, the original clean data, and not $\widetilde{\textbf{x}}$, the noisy inputs.
The loss function minimized by stochastic gradient descent therefore becomes:
\begin{equation}
  \dfrac{1}{n}\sum\limits_{i=1}^n L_{q_{D}(\widetilde{\textbf{x}}|\textbf{x})}(\textbf{x}^{(i)},g_{\theta'}(f_{\theta}(\tilde{\textbf{x}}^{i}))),
\end{equation}
where all the notations have the same meaning as in Equation 7.

\subsection{CNN-LSTM Model Architecture}
\noindent The architecture of the CNN-LSTM model is shown in Table 1.

\begin{table}
\caption{\label{tab:table2}%
Hyper-parameters of the constructed CNN-LSTM de-noising autoencoder architecture used for this paper (See text in Section II A). \\
}
\begin{ruledtabular}
\begin{tabular}{ccc}
\textrm{Layer}&
\textrm{Output Shape}&
\textrm{Kernels/Neurons} \\ [0.5ex]
\hline\hline
Input & (516, 4, 1) & -\\
\hline
Conv1D  & \multirow{2}{5em}{(516, 4, 32)} & 32\\
(Time-distributed) &  & \\
\hline
MaxPool1D & \multirow{2}{5em}{(516, 2, 32)} & -\\

(Time-distributed) &  & \\
\hline
Conv1D & \multirow{2}{5em}{(516, 2, 16)}  & 16\\

(Time-distributed) &  & \\
\hline
Flatten & {(516, 32)}  & -\\
\hline
LSTM  & \multirow{2}{5em}{(516, 200)}  & 100\\

(Bi-directional)  &  & \\
\hline
LSTM  & \multirow{2}{5em}{(516, 200)}  & 100\\

(Bi-directional)  &  & \\
\hline
LSTM  & \multirow{2}{5em}{(516, 200)}  & 100\\

(Bi-directional)  &  & \\
\hline
Dense  & \multirow{2}{4em}{(516, 1)}  & 1\\

(Time-distributed)  &  & \\
\hline

\end{tabular}
\end{ruledtabular}
\end{table}

\end{document}